\documentclass[aps,twocolumn,floats,showpacs,superscriptaddress,prb]{revtex4}
\usepackage{amsmath,amssymb,graphicx,epsf,epsfig}

\begin{document}

\title{The Nernst effect from fluctuating pairs in the pseudogap phase}
\author{Alex Levchenko}
\affiliation{Materials Science Division, Argonne National Laboratory, Argonne IL 60439}
\author{M. R. Norman}
\affiliation{Materials Science Division, Argonne National Laboratory, Argonne IL 60439}
\author{A. A. Varlamov}
\affiliation{Materials Science Division, Argonne National Laboratory, Argonne IL 60439}
\affiliation{SPIN-CNR - Viale del Politecnico 1, I-00133 Rome, Italy}

\begin{abstract}
The observation of a large Nernst signal in cuprates above the
superconducting transition temperature has attracted much attention.
A potential explanation is that it originates from superconducting fluctuations.
Although the Nernst signal is indeed consistent with gaussian fluctuations for overdoped
cuprates, gaussian theory fails to describe the temperature dependence seen
for underdoped cuprates. Here, we consider the vertex correction to gaussian
theory resulting from the pseudogap. This yields a Nernst signal in good
agreement with the data.
\end{abstract}

\date{September 12, 2010}
 \maketitle

For most of the doping phase diagram, high temperature
superconductivity in the cuprates emerges from a normal state where
an energy gap is already present.\cite{Timusk} The origin of this
`pseudogap' is the subject of much debate.\cite{Kallin} One of the
most interesting observations in the pseudogap phase is the
existence of a large Nernst signal.\cite{Xu,Wang} The Nernst effect
is the generation of a transverse electric field by
a thermal gradient in the presence of a magnetic field perpendicular
to both. Since vortices carry entropy,
it is natural to attribute such a large Nernst signal in proximity
to the superconducting transition temperature $T_{c}$
to vortex-like excitations.\cite{Xu,Wang}  Moreover, invoking vortices is
consistent with the Nernst signal smoothly going through $T_{c}$. On the other
hand, it is not clear whether vortices give an adequate
description of the physics of fluctuating superconductors except
very near $T_{c}$.\cite{Uss04} The free vortex density above the
Kosterlitz-Thouless temperature increases exponentially with
temperature, a dependence which is inconsistent with the near power-law
decrease of the actual Nernst signal above $T_{c}$. Moreover,
the recent observation of a \textit{negative} Nernst signal for
underdoped YBa$_{2}$Cu$_{3}$O$_{6+y}$ has further complicated the
story.\cite{Louis} This negative signal has been argued to be a
consequence of density wave reconstruction of the Fermi surface.

Here, we take the point of view that the dominant contribution to
the Nernst signal in the pseudogap phase is indeed due to
fluctuating pairs. This is supported by the close correspondence of
the Nernst signal with the fluctuational diagmagnetism.\cite{Li10}
On the other hand, we note that although existing theories based on
Ginzburg-Landau or diagrammatic approaches
\cite{Dorsey,Uss02,Serbyn} give a good description of the Nernst
data for overdoped cuprates, they do less well for underdoped
cuprates. We attribute this to the presence of the pseudogap.

As discussed in Ref.~\onlinecite{Serbyn}, it is the direct
contribution from fluctuating pairs - the Aslamazov-Larkin (AL)
contribution \cite{AL} - which governs the Nernst signal over a wide
range of temperatures above $T_c$, so we focus on that. The AL
contribution to the Nernst coefficient is obtained from the electric
current-heat current Kubo response kernel
$\Lambda_{xy}$.\cite{Uss02,Serbyn} The latter can be expressed in
terms of corresponding electric and heat vertex blocks (triangular
graphs) connected by interaction lines (pair fluctuations). The
vertex block can be expressed as
$\mathrm{Tr}\{\gamma^{(e,h)}\mathbf{B}\}$ with
$\mathbf{B}=\left\langle \mathbf{v}GGG\right\rangle$ where
$\mathbf{v}$ is the Fermi velocity, and $\langle...\rangle$
indicates disorder averaging. The factor $\gamma$ differentiates the
electric vertex ($\gamma^{(e)}=e$) from the heat vertex
$\gamma^{(h)}$, which we will discuss below. Disorder averaging in
$\mathbf{B}$ leads to the presence of two Cooperons and the
renormalization of the free electron Greens function $G$ by impurity
scattering. In order to account for the pseudogap, we replace this
$G$ which was previously used to compute this block by the broadened
BCS Greens function
\begin{equation}
G(k,\varepsilon )=-\frac{i\bar{\varepsilon}+\xi _{k}}{\bar{\varepsilon}%
^{2}+\xi _{k}^{2}+\Delta _{k}^{2}}  \label{gbroad}
\end{equation}
as this gives a good description of photoemission data in the
pseudogap phase.\cite{Norm98,Tallon,Norm07} In Eq.~\eqref{gbroad},
$\Delta _{k}$ is the momentum dependent pseudogap and
$\bar{\varepsilon}=\varepsilon +\Gamma \,\mathrm{sgn}(\varepsilon )$
with $\Gamma $ the scattering rate. By recomputing the
electromagnetic vertex block with this $G$, we find that $\mathbf{B}$ is
renormalized by a function of $\Delta /\Gamma$ where $\Delta$ is the
maximum value of the pseudogap.
Assuming a $T$ independent $\Delta$ and $\Gamma\sim T$ as
observed in photoemission,\cite{Norm98,Kanigel} this renormalization
results in a fluctuation Nernst signal which drops off considerably faster with
temperature than the gaussian result. As we show, this gives a good
description of the Nernst data for underdoped cuprates.

We assume the standard expression for the pair propagator whose retarded
component has the form \cite{book}
\begin{equation}
L_{R}(q,\omega )=-\frac{1}{N_{0}}\frac{1}{\epsilon -i\pi \omega
/8T+\eta q^{2}}  \label{L-gaussian}
\end{equation}
Here, $\epsilon =\ln (T/T_{c})$, $N_{0}$ is the density of states,
and $\eta =\pi D/8T $ where $D$ is the diffusion constant. The
Nernst coefficient can be
expressed in terms of the electrical ($\hat{\sigma}$) and thermoelectric $(%
\hat{\alpha})$ tensors as $\nu =(\alpha _{xy}\sigma _{xx}-\alpha _{xx}\sigma
_{xy})/H(\sigma _{xx}^{2}+\sigma _{xy}^{2})\approx \alpha _{xy}/H\sigma
_{xx} $. The second (approximate) expression, which becomes exact if
particle-hole symmetry is present, gives a good approximation to $\nu $
for the case of superconducting fluctuations since $\alpha _{xy}\gg \alpha
_{xx}$ and $\sigma _{xx}\gg \sigma _{xy}$ (see Ref.~\onlinecite{book} for a
corresponding discussion). The transverse thermoelectric coefficient, $%
\alpha _{xy}=\bar{\alpha}_{xy}+cM_{z}/T$, consists of two
independent contributions: the response of the total current to the
applied electric and magnetic fields ($\bar{\alpha}_{xy}$), and the
magnetization currents as derived from the equilibrium magnetization
$M_{z}$.  Like Refs.~\onlinecite{Uss02,Serbyn}, we focus on the first
contribution
\begin{equation}
\bar{\alpha}_{xy}=\frac{H}{cT}\lim_{Q,\Omega\to0}\frac{1}{Q\Omega}
\mathrm{Re}[\Lambda_{xy}^{R}(Q,\Omega)]
\end{equation}
assuming here weak field limit. The electric current-heat current
Kubo response kernel
\begin{eqnarray}
&&\hskip-.25cm\Lambda_{xy}(Q,i\Omega_{m})=-4e^2T\sum_{q,\omega
_{n}}B_{x}(q)B^2_{y}(q)(i\omega_n+i\Omega_m/2)
\notag  \label{Qxy} \\
&&
\left[L(q-Q_x,i\omega_{n})L(q,i\omega_n)L(q,i\omega_{n}+i\Omega_{m})\right.\nonumber\\
&&+\left.L(q,i\omega_n)L(q,i\omega_n+i\Omega_m)L(q+Q_x,i\omega_n+i\Omega_m)\right]
\end{eqnarray}
is written in Matsubara representation, where we have assumed that
the heat vertex is $-i\omega_n/(2e)$ times the electric
vertex.\cite{heat} In Eq.~\eqref{Qxy}, the Greens function block
$B$, whose renormalization is the subject of this paper, is assumed
to be independent of frequency. This approximation is formally exact
in the immediate vicinity of the transition temperature.
Nevertheless, a rigorous extension of this approximation to a wider
range of temperatures above $T_{c}$ demonstrates that the
Ginzburg-Landau result, $\nu^{AL}\sim \left( T-T_{c}\right) ^{-1}$,
remains valid even far from $T_{c}$ if one substitutes $T-T_{c}$ by
the more general expression $T\ln(T/T_c)$. We note that in the
gaussian approximation \cite{book}
\begin{equation}
B_{i}^{G}(q)=-2N_{0}\eta q_{i}
\end{equation}
For the following linear response calculation we expand the
propagators in Eq.~\eqref{Qxy} to the leading order in $Q$, namely
$L(q\pm Q_x,i\omega_n)\to\pm Q\partial_{q_x}L(q,i\omega_n)$, and
noting that
\begin{equation}
\frac{\partial L(q,i\omega_n )}{\partial
q_{x}}=-B_{x}^{G}(q)L^{2}(q,i\omega_n )
\end{equation}
we thus find from Eq.~\eqref{Qxy}
\begin{eqnarray}
&&\hskip-.5cm\Lambda_{xy}(Q,i\Omega _{m})=\\
&&\hskip-.5cm-4e^{2}QT\sum_{q,\omega_n}
B_{x}(q)B_{x}^{G}(q)B_{y}^{2}(q)(i\omega_n+i\Omega_m/2)\nonumber\\
&&\hskip-.5cm\big[L^{3}(q,i\omega_{n})L(q,i\omega_{n}+i\Omega_{m})
-L(q,i\omega_{n})L^{3}(q,i\omega_{n}+i\Omega_{m})\big]\nonumber
\end{eqnarray}
Performing the summation over the Matsubara frequency $\omega _{n}$
by using contour integration,
$T\sum_{\omega_n}f(i\omega_n)=\frac{1}{4\pi
i}\oint_Cd\omega\coth\frac{\omega}{2T}f(\omega)$ with two
branch-cuts at $\mathrm{Im}(\omega)=0$ and
$\mathrm{Im}(\omega)=-\Omega_m$, followed by an analytic
continuation $\Omega_{m}\rightarrow -i\Omega $, and keeping only the
linear in $\Omega $ contribution from $\Lambda_{xy}^{R}(Q,\Omega)$,
one finds for the transverse thermoelectric coefficient
\begin{eqnarray}
&&\hskip-.75cm \bar{\alpha}_{xy}^{AL}=\frac{4e^{2}H}{\pi cT}
\sum_{q}B_{x}^{G}(q)B_{x}(q)B_{y}^{2}(q)\int_{-\infty
}^{+\infty}d\omega
\coth \frac{\omega }{2T}  \notag \\
&&\hskip-.75cm \left\{\!\lbrack
\mathrm{Re}L_{R}(q,\omega)]^{3}\mathrm{Im}L_{R}(q,\omega )\!+\!
\mathrm{Re}L_{R}(q,\omega )[\mathrm{Im}L_{R}(q,\omega )]^{3}\right\}
\end{eqnarray}
After the remaining momentum and energy integrations, we get the
result (restoring $\hbar $)
\begin{equation}
\bar{\alpha}_{xy}^{AL}=\frac{e}{2\pi \hbar }\frac{\xi^{2}_{GL}}{\ell
_{H}^{2}} (B/B^{G})^{3}
\end{equation}
where $\ell _{H}=\sqrt{\hbar c/eH}$ is the magnetic length and
$\xi^{2}_{GL}=\eta /\ln (T/T_{c})$. For $B=B^{G}$, this is the well
known expression for the Nernst effect from fluctuating
pairs.\cite{Uss02,Serbyn}

At this point, all we have done is to rederive the gaussian
expression. The reason we have done this is to demonstrate
explicitly where the $B_{i}$ current vertices enter. We now discuss
the renormalization of $B_{i}$ due to the pseudogap. The expression
for the vertex is \cite{AL}
\begin{equation}
B_{i}(q,\omega ,\Omega )=T\sum_{p,\varepsilon }v_{i}(p)G(p,\varepsilon
)G(p,\varepsilon +\Omega )G(q-p,\omega -\varepsilon )
\end{equation}
where $\varepsilon $ is the fermionic loop frequency, $\omega $ the
bosonic frequency that enters the fluctuation propagator, $\Omega $
the external field frequency (set to zero for the dc response), and
$G$ the Greens function.\cite{cooperons}
We will now recalculate $B_{i}$ using the
pseudogap Greens function. Substituting Eq.~\eqref{gbroad} in the
previous equation, taking the dc limit, and approximating
$G(q-p,\omega -\varepsilon )\approx G(q-p,-\varepsilon )$ we obtain
\begin{equation}
B_{i}(q)\simeq -T\sum_{p,\varepsilon }v_{i}\left( \frac{i\bar{\varepsilon}%
+\xi _{p}}{\bar{\varepsilon}^{2}+\xi _{p}^{2}+\Delta _{p}^{2}}\right) ^{2}%
\frac{-i\bar{\varepsilon}+\xi _{q-p}}{\bar{\varepsilon}^{2}+\xi
_{q-p}^{2}+\Delta _{p}^{2}}
\end{equation}
Keeping the term to linear order in $q$,
\begin{eqnarray}
B_{i}(q) &\simeq &\frac{1}{2}Tv_{F}^{2}q_{i}\sum_{p,\varepsilon }\left(
\frac{i\bar{\varepsilon}+\xi _{p}}{\bar{\varepsilon}^{2}+\xi _{p}^{2}+\Delta
_{p}^{2}}\right) ^{2}  \notag \\
&&\left[ \frac{1}{\bar{\varepsilon}^{2}+\xi _{p}^{2}+\Delta _{p}^{2}}-\frac{%
2\xi _{p}(-i\bar{\varepsilon}+\xi _{p})}{(\bar{\varepsilon}^{2}+\xi
_{p}^{2}+\Delta _{p}^{2})^{2}}\right]
\end{eqnarray}
Converting the $p$ sum to an integral, we have
\begin{eqnarray}
B_{i}(q) &\simeq &\frac{1}{2}N_{0}Tv_{F}^{2}q_{i}\sum_{\varepsilon }\int
\frac{d\vartheta }{2\pi }\int_{-\infty }^{+\infty }d\xi   \notag \\
&&\left[ \frac{\xi
^{2}-\bar{\varepsilon}^{2}}{(\bar{\varepsilon}^{2}+\xi ^{2}+\Delta
_{\vartheta }^{2})^{3}}-\frac{2\xi ^{2}(\xi ^{2}+\bar{\varepsilon
}^{2})}{(\bar{\varepsilon}^{2}+\xi ^{2}+\Delta _{\vartheta
}^{2})^{4}}\right]
\end{eqnarray}
where $\Delta _{\vartheta }=\Delta \cos 2\vartheta $ (i.e., we assume a
d-wave pseudogap). Next we perform the $\xi $ integral by introducing $\xi
=\mu x$ with $\mu =\sqrt{\bar{\varepsilon}^{2}+\Delta _{\vartheta }^{2}}$
\begin{eqnarray}
B_{i}(q) &\simeq &\frac{1}{4\pi }N_{0}Tv_{F}^{2}q_{i}\sum_{\varepsilon }\int
\frac{d\vartheta }{(\bar{\varepsilon}^{2}+\Delta _{\vartheta }^{2})^{5/2}}
\notag \\
&&\int_{-\infty }^{+\infty }\!\!dx\left[ \frac{\mu ^{2}x^{2}-\bar{\varepsilon}%
^{2}}{(1+x^{2})^{3}}-\frac{2x^{2}(\mu ^{2}x^{2}+\bar{\varepsilon}^{2})}{%
(1+x^{2})^{4}}\right]
\end{eqnarray}
The $x$ integral is trivial, and we find
\begin{equation}
B_{i}(q)\simeq -\frac{1}{8}N_{0}Tv_{F}^{2}q_{i}\sum_{\varepsilon }\int \frac{%
\bar{\varepsilon}^{2}d\vartheta }{(\bar{\varepsilon}^{2}+\Delta _{\vartheta
}^{2})^{5/2}}
\end{equation}
The angular integral is easily performed, leading to
\begin{eqnarray}
B_{i}(q) &=&-\frac{1}{3}N_{0}Tv_{F}^{2}q_{i}\sum_{n=0}^{\infty }\frac{1}{%
\bar{\varepsilon}^{2}}\frac{1}{(\bar{\varepsilon}^{2}+\Delta ^{2})^{3/2}}
\notag \\
&&\left[ (4\bar{\varepsilon}^{2}+2\Delta ^{2})E(\lambda )-\bar{\varepsilon}%
^{2}K(\lambda )\right]
\end{eqnarray}
where $\lambda =\Delta /\sqrt{\bar{\varepsilon}^{2}+\Delta ^{2}}$
and $K$ and $E$ are elliptic functions. As the sum converges at $n
\gtrsim \Delta /T$, one can approximate the elliptic functions by
their value at zero argument, which is $\pi /2$. We will also take
the `zero T' limit by converting the Matsubara sum to an integral,
noting that the dominant $T$ dependence comes from the $T$
dependence of $\Gamma $. This gives
\begin{equation}
B_{i}(q)=-\frac{N_{0}v_{F}^{2}}{12}q_{i}\int_{0}^{\infty }\frac{d\varepsilon
}{\bar{\varepsilon}^{2}}\frac{3\bar{\varepsilon}^{2}+2\Delta ^{2}}{(\bar{%
\varepsilon}^{2}+\Delta ^{2})^{3/2}}
\end{equation}
and after the remaining integration results in
\begin{equation}
B_{i}(q)=-\frac{\pi ^{2}}{12}N_{0}\xi _{0}^{2}q_{i}\left[
\frac{(\Gamma /\Delta )^{2}+2}{(\Gamma /\Delta )\sqrt{(\Gamma
/\Delta )^{2}+1}}-1\right] \label{B}
\end{equation}
where we have exploited the BCS relation $\xi _{0}=\hbar v_{F}/\pi \Delta $.
Inserting the gaussian expression for $B$, we obtain
\begin{equation}
B/B^{G}=\frac{\pi ^{2}}{24}\frac{\xi _{0}^{2}}{\eta }\left[
\frac{(\Gamma /\Delta )^{2}+2}{(\Gamma /\Delta )\sqrt{(\Gamma
/\Delta )^{2}+1}}-1\right]
\end{equation}

\begin{figure}[tbp]
\centerline{\includegraphics[width=6cm]{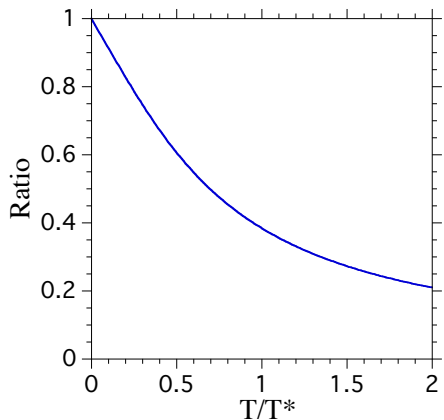}} \caption{Ratio of the
renormalized $B$ vertex to its small $\protect\Gamma$
limit versus $T/T^*$ where $T^*$ is the pseudogap temperature, and $\protect%
\Gamma/\Delta=\protect\sqrt{3}T/T^*$ with the coefficient chosen so
that the spectral gap disappears at $T^*$.} \label{fig1}
\end{figure}

In the limit of small $\Gamma \ll \Delta $, the term in parenthesis reduces to
$2\Delta /\Gamma $. Since $\Gamma \sim T$ and $\eta \sim 1/T$, then
in this limit, the ratio is a constant, and one obtains the same
functional form for the Nernst as in the gaussian approximation. On
the other hand, as $\Gamma $ increases, the ratio decreases from
unity as can be seen in Fig.~1. This leads to a Nernst signal which
decays more rapidly than the gaussian result, since three
renormalized $B$ vertices enter the expression for the Nernst:
\begin{equation}
\bar{\alpha}_{xy}=\bar{\alpha}_{xy}^{G}(B/B^{G})^{3}
\end{equation}

We now consider the Nernst data for La$_{2-x}$Sr$_{x}$CuO$_{4}$.\cite{Wang}
The advantage of these data is that after the normal carrier
background has been subtracted, the Nernst signal is positive,
 and therefore complications due to density
wave reconstruction can to first order be ignored. In Fig.~2, we
show the Nernst signal, $\nu =-E_{y}/(H_z \nabla_x T)$, for four different
dopings. For the overdoped compound, the gaussian expression for
$\bar{\alpha}_{xy}$ fits the data quite well, but not for the three
underdoped compounds where the pseudogap is present. Instead, we find
that the corrected expression provides a good description of the data (with
pseudogap temperatures $T^*$ ranging from about 200 to 300 K).

\begin{figure}[tbp]
\centerline{\includegraphics[width=8cm]{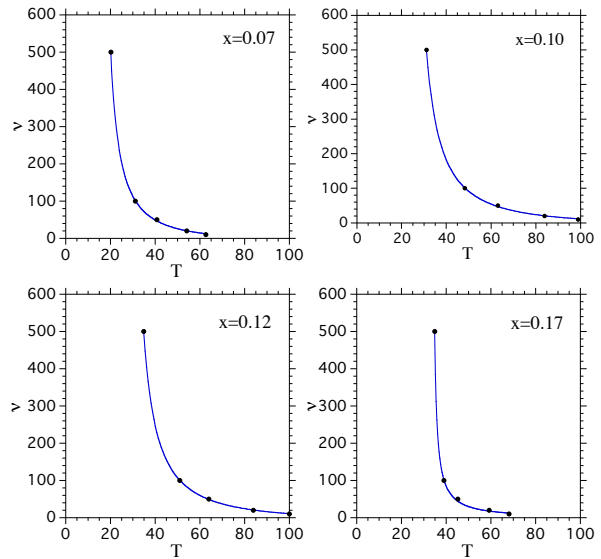}} \caption{Nernst
signal, $\protect\nu$, versus $T$ for La$_{2-x}$Sr$_x$CuO$_4$ for
four different values of the doping, $x$.\protect\cite{Wang} The
curve
for x=0.17 is a fit to the gaussian expression for $\bar{\protect\alpha}%
_{xy} $. The other curves include the vertex correction as described in this
paper. }
\label{fig2}
\end{figure}

We now turn to a brief discussion of the bosonic contribution to the
conductivity.  As with the Nernst, the paraconductivity observed in
underdoped cuprates well above $T_c$ falls off like $T^{-\delta }$
with $\delta \approx 3$.\cite{Leridon} We recall that when extended
to the higher temperature regime, gaussian theory (in 2D) predicts
for the Aslamazov-Larkin \cite{RVV} contribution to the conductivity
$\sigma _{xx}^{AL}=e^{2}/8\pi ^{2}\ln ^{3}(T/T_{c})$ while for the
Maki-Thompson \cite{AV-ARV} contribution,
$\sigma_{xx}^{MT}=\pi^{2}e^{2}\ln
(1/\gamma_{\varphi})/192\ln^{2}(T/T_{c})$ where $\gamma _{\varphi }$
is the dephasing rate. In either case, the $\ln ^{-n}(T/T_{c})$
decay is too slow to explain the experimental data. We now argue
that the same vertex renormalization can account for the faster
decay of the fluctuational conductivity. We start from the
definition of the conductivity in the linear response regime $\sigma
_{xx}^{AL}=-\mathrm{Im}[\Lambda_{xx}^{R}(\Omega )]/\Omega $ where
the current-current response kernel is
\begin{equation}
\Lambda_{xx}(i\Omega _{m})=-4e^{2}T\sum_{q,\omega
_{n}}B_{x}^{2}(q)L(q,i\omega _{n})L(q,i\omega _{n}+i\Omega _{m})
\end{equation}
After Matsubara summation and analytic continuation, this
reduces to
\begin{eqnarray}
\Lambda_{xx}^{R}(\Omega ) &=&-\frac{2e^{2}}{\pi
}\sum_{q}B_{x}^{2}(q)\int_{-\infty }^{+\infty }d\omega \coth
\frac{\omega }{2T}\mathrm{Im}L_{R}(q,\omega )
\notag \\
&&\times \lbrack L_{R}(q,\omega +\Omega )+L_{A}(q,\omega -\Omega )]\qquad
\end{eqnarray}
In the dc limit, we expand $L_{R}(q,\omega +\Omega )+L_{A}(q,\omega -\Omega
)\simeq 2i\Omega \partial _{\omega }\mathrm{Im}L_{R}(q,\omega )$, integrate
by parts over $\omega $, and as a result find for the AL contribution to the conductivity
\begin{equation}
\sigma _{xx}^{AL}=\frac{e^{2}}{\pi T}\sum_{q}B_{x}^{2}(q)\int_{-\infty
}^{+\infty }\frac{d\omega }{\sinh ^{2}\frac{\omega }{2T}}[\mathrm{Im}%
L_{R}(q,\omega )]^{2}
\end{equation}
This is formally the same expression as in gaussian theory except for
the vertex function now defined by Eq.~\eqref{B}. We thus find as a result
(restoring $\hbar $)
\begin{equation}
\sigma _{xx}^{AL}=\frac{e^{2}}{16\hbar \ln(T/T_{c})}(B/B^{G})^{2}
\end{equation}
where the renormalization factor $(B/B^{G})^{2}$ provides a faster
power-law decay, consistent with the data.\cite{Leridon}

We would like to conclude with the observation that although we assumed a `BCS'
expression for the pseudogap Greens function, $G$, any theory of
the pseudogap with a $T$ independent d-wave like gap and a scattering rate
proportional to $T$ will yield results equivalent  to those derived here.
Similar conclusions have been reached in regards to the fermionic contribution
to various transport properties in the pseudogap phase.\cite{Alex10}

In summary, we note that the Nernst signal and fluctuational
conductivity for \textit{underdoped} compounds drops more rapidly
with temperature than predicted from a gaussian theory of
fluctuating pairs. This discrepancy is nicely accounted for by a
vertex correction to the fermionic current block due to the
pseudogap.

This work was supported by the U.~S. DOE, Office of Science, under contract
DE-AC02-06CH11357.  The authors acknowledge helpful discussions with
M.~N.~Serbyn.


\begin{thebibliography}{99}
\bibitem{heat}
Here, we have used the same heat vertex as Ussishkin \textit{et
al.}\cite{Uss02} so as reproduce the phenomenological
Ginzburg-Landau result for the Nernst.  The actual heat vertex
should be twice as large as discussed by M. Yu. Reizer and A. V.
Sergeev, Phys. Rev. B {\bf 50}, 9344 (1994).  The latter is
supported by the detailed calculations of Ref.~\onlinecite{Serbyn}
where the full frequency dependence of the heat vertex was kept.
\bibitem{cooperons} In writing this expression, we have made use of the fact
that since the characteristic scale of the bosonic momentum is $q \ll k_F$,
we can set $q \rightarrow 0$ in the Cooperons, $C$, thus approximating
$C(q,\varepsilon+\Omega,\omega-\varepsilon)C(q,\varepsilon,\omega-\varepsilon)\approx
C^2(0,\varepsilon,-\varepsilon)\approx1$.
\end{thebibliography}
\end{document}